\documentclass[final,runningheads]{svjour2}%
\usepackage{graphicx}
\usepackage{mathptmx}
\usepackage{amsmath}
\usepackage{bm}
\usepackage{amsfonts}
\usepackage{amssymb}%

\smartqed
\journalname{\normalsize J Stat Phys (2007) 128: 1321-1336}
\begin{document}

\title{\Large Thermostats~for~``slow''~configurational~modes}

\author{Alex~A.~Samoletov
\and Carl~P.~Dettmann
\and Mark~A.~J.~Chaplain }

\institute{A.A. Samoletov $\boxtimes$ \and M.A.J. Chaplain
\at Division of Mathematics, University of Dundee, Dundee DD1 4HN, UK \\
\email{asamolet@maths.dundee.ac.uk } \\
\\
M.A.J. Chaplain \\
\email{chaplain@maths.dundee.ac.uk}\\
\\
A.A. Samoletov $\cdot$ C.P.~Dettmann
\at Department of Mathematics, University of Bristol, Bristol BS8 1TW, UK \\
\\
C.P. Dettmann \\
\email{Carl.Dettmann@bristol.ac.uk}\\
\\
\emph{Present address:} \\
A.A. Samoletov
\at Institute for Physics and Technology, NASU, Donetsk 83114, Ukraine \\
\email{samolet@kinetic.ac.donetsk.ua}
}

\titlerunning{J Stat Phys (2007) 128: 1321-1336}
\authorrunning{J Stat Phys (2007) 128: 1321-1336}

\date{Published online: 26 July 2007}

\maketitle

\begin{abstract}
Thermostats are dynamical equations used to model thermodynamic variables such
as temperature and pressure in molecular simulations. For computationally
intensive problems such as the simulation of biomolecules, we propose to
average over fast momentum degrees of freedom and construct thermostat
equations in configuration space. The equations of motion are
\emph{deterministic} analogues of the Smoluchowski dynamics in the method of
stochastic differential equations.
\end{abstract}

\section{Introduction}
\label{intro}

One of the most ambitious challenges in mathematical modelling of
biological processes is to describe dynamics of two major biological
events within a cell - DNA molecule replication and transcription.
In both cases, dynamical properties, especially large amplitude
conformational changes of the double-stranded DNA molecule play a
vital part. The principal physical feature of biological functioning
of biomolecules is that they operate at ambient physiological
temperature, pressure and solvent conditions. Thus, the surrounding
physiological solvent plays the role of a thermostat, among others.
To properly thermostat this dynamics, especially on a biological
(``slow'') time scale, we propose a suitable and effective
temperature control of interest both for the practice of numerical
simulation and the general theory of dynamical systems.

For a recent and comprehensive review of the problem outlined above
we refer to \cite{mp04}. Current approaches commonly use the
Nos\'{e}-Hoover thermostat method (for review see
\cite{T1,T2,T3,T4,T5}). This canonical thermostat method involves
integration of both position and momentum phase space variables.
However, for problems that are related to slow conformational
changes of biomolecules, integration of fast momentum variables
appears superfluous from a theoretical point of view (unobservable
variables) as well expensive in the sense of numerical simulation.

In this paper, a novel approach to the problem of slow
conformational changes in thermostatting dynamics is presented. The
method is based on an analogy with the derivation of the
Smoluchowski dynamics~(Eq.~(\ref{Smoluchowski}) below) from the
Langevin stochastic dynamics~(Eq.~(\ref{Langevin}) below)
\cite{Gard04,kramers40,as99}. The configurational deterministic
thermostat is constructed so as to effect the temperature control
via certain dynamics of the relaxation rate variable and also
involves a dynamically fluctuated collective force to ensure the
ergodicity property. As the result, the temperature control is
connected to the specific configurational temperature recently
introduced in \cite{rugh97,evans00} in a different context. The new
configurational thermostat can be combined with complementary
temperature control via a dynamically fluctuated virial function
\cite{hamilton90,sc05} that helps to enhance the efficiency of the
thermostat and, more importantly because biological molecules are
functioning at a constant pressure, implements the pressure control
into the corresponding dynamics. Moreover, the configurational
thermostat admits stimulation by a chain of thermostats similar to
Nos\'{e}-Hoover \cite{chain} and a stochastically driven method (see
below). To test the new configurational thermostat, corresponding
simulations of a one-dimensional harmonic and the Morse oscillator
dynamics are given, providing a stringent test of the ergodicity
property.

Remark: Braga and Travis \cite{BT05} have recently proposed a
thermostat based on the Smoluchowski equation and configurational
temperature ideas (but retaining the momentum variables), citing an
early preprint of this paper \cite{SCD04}.

\section{Preliminaries}
\label{sec:2}%
Section \ref{sec:2.1} briefly reviews the Langevin and Smoluchowski
approaches to a system in contact with the environment. Section
\ref{sec:2.2} presents the Nos\'{e}-Hoover thermostat scheme in
brief in the form of consequent steps and logic that we follow in
Section \ref{sec:3}. Section \ref{sec:2.3} presents the virial
thermostat scheme. Section \ref{sec:2.4} presents a stochastically
stimulated thermostat scheme in the context of Nose-Hoover dynamics.
Following these preliminary sections our proposed configurational
thermostats are presented in Sections~\ref{sec:3} and \ref{sec:4}
and tested numerically in Section~\ref{sec:5}. Section~\ref{sec:6}
concludes.

\subsection{Stochastic dynamics}
\label{sec:2.1}

The early and successful attempts to describe the dynamics of a
mechanical system being in contact with an environment playing a
role of the thermostat are based on the concept of the stochastic
differential equation \cite{Gard04}. Langevin's equation for the
case of one-dimensional motion of a particle mass $m$ in a potential
field $V(q)$ provides the characteristic example,
\begin{equation}
m\dot{q}=p,\quad\dot{p}=-\nabla V(q)-\gamma p+\sqrt{2D}f(t),
\label{Langevin}%
\end{equation}
where the friction coefficient $\gamma$ and the intensity $D$ of the external
random force $f(t)$ are connected by the relation, $D=m\gamma k_{\mathrm{B}}%
T$; here $f(t)$ is the generalized Gaussian stochastic process,
\textquotedblleft white noise\textquotedblright, with characteristic
cumulants $\left\langle f(t)\right\rangle =0$ and $\left\langle
f(t)f(t^{\prime })\right\rangle =\delta(t-t^{\prime})$. The
equilibrium solution of the corresponding Fokker-Planck equation ,
\[
\frac{\partial \rho}{\partial t}= -\frac{p}{m} \frac{\partial
\rho}{\partial q}+\frac{\partial}{\partial p}\left[ \left( \frac{d
V}{d q} + \gamma p \right) \rho \right] +D\frac{{\partial}^2
\rho}{\partial p^2},
\]
 is known \cite{Gard04}, $\rho_{\infty }\propto\exp\left[  -\left(
p^{2}/2m+V(q)\right)  /(k_{\mathrm{B}}T)\right] $.

Langevin's equation (\ref{Langevin}) is a prototype of the Nos\'{e}-Hoover
deterministic dynamics (after generalisation of the last with the term
$\vec{\xi}$ as in Eq.~(\ref{VSdyn}) it is especially evident). But while the
Langevin dynamics generate all the sample trajectories and the corresponding
measure, the Nos\'{e}-Hoover dynamics produce a single sample trajectory with
the correct canonical ensemble statistics when ergodicity holds.

Often details of the dynamics of a system on short time scales are not needed
for a dynamical description of the observable variables. However, then the
Smoluchowski limit of (\ref{Langevin}),
\begin{equation}
\label{Smoluchowski}m\dot{q}=-\tau\nabla V(q)+\sqrt{2mk_{\mathrm{B}}T\tau
}f(t),
\end{equation}
where $\tau=\gamma^{-1}$, is an appropriate formulation
\cite{Gard04,kramers40,as99}. Only the position variable is involved
in this equation. Formally, it is supposed that the momentum
variable relaxes to the local equilibrium state. The corresponding
Smoluchowski equation has the Boltzmann distribution as the
equilibrium solution \cite{Gard04}. Eq.~(\ref{Smoluchowski}),
without the random perturbation, appears as a dissipative dynamics
with $V(q)$ playing a role of the Lyapunov function,
$\dot{V}=-\tau\left(  \nabla V(q)\right)  ^{2}\leq0$. Thus, the full
dynamics is a superposition of relaxation to a minimum of the
potential and random perturbations that occasionally expel the
system outside the vicinity of the minimum. This process
equilibrates the system.

\subsection{The Nos\'{e}--Hoover dynamics}

\label{sec:2.2}

Consider a system of $N$ particles of mass $m=\{m_{k}\}_{k=1}^{N}$
with vectors of coordinate $\bm{q}=\{{\mathbf{q}}_{k}\}_{k=1}^{N}$
and momentum $\bm{p}=\{{\mathbf{p}}_{k}\}_{k=1}^{N}$, and the
Hamiltonian function
$H(\bm{q},\bm{p})=\sum{\mathbf{p}^{2}/(2m)}+V(\bm{q})$, where
$V(\bm{q})$ is the potential function of the system of particles.
Here and in what follows $\sum$ means sum
over all particles of the system (\textit{e.g. }$\sum{\mathbf{p}%
^{2}/(2m)\equiv}\sum_{k=1}^{N}{\mathbf{p}_{k}^{2}/(2m}_{k}{)}$).
Physical space is $\mathrm{d}$-dimensional, $\mathrm{d}=1,2,3$, and
respectively the system has $\mathrm{d}N$ degrees of freedom and
$2\mathrm{d}N$-dimensional phase space $\mathcal{M}=( \bm{q},\bm{p})
$. To simulate the canonical statistics with deterministic dynamics
we are in need of non-canonical forces to cool/heat the system and
equilibrate the dynamics. The Nos\'{e}-Hoover scheme
\cite{T1,T2,T3,T4,T5} is characterised by the simplest form of
non-canonical forces, the Rayleigh friction, $-\zeta\mathbf{p}$, and
supposes the Gaussian fluctuation of the single thermostat variable
$\zeta$ at the equilibrium.

Let us emphasise here the following scheme we regularly utilise
throughout the paper,
\begin{itemize}
 \item first, we consider a deformation of the dynamical equations
 in the phase space $\mathcal M$
 with a constant parameter (the parameter $\zeta$ in this Section).
 Even if it is obvious that with the constant deformation parameter
 the equations of motion are too simple to generate the canonical
 statistics they are a convenient starting point for further
 discussion;
 \item then, requiring that the Liouville equation corresponding to
 these deformed equations of motion has the canonical distribution
 as a steady state solution we arrive at a condition involving the
 temperature into dynamics (the condition (\ref{c1}) in this
 Section). Under this condition on the dynamic variables the
 deformation parameter becomes variable (nonconstant). This explicit
 contradiction with original supposition concerning the deformation
 parameter results in conclusion that the deformed equations of
 motion together with the obtained temperature control condition
 cannot generate the canonical statistics;
 \item finally, to preserve in essential the structure of the Liouville equation
corresponding to the dynamical equations with constant deformation
parameter and thus allow generation of the canonical statistics, we
extend the phase space $\mathcal M $ and consider the deformation
parameter as an independent variable with its own equation of
motion. This results in a dynamical temperature control.
\end{itemize}

Firstly we consider $\zeta$\ as a constant, $\zeta\neq0$, but
further it will be endowed with its own equation of motion.
Dynamical equations take the form,
\begin{equation}
\mathbf{\dot{q}}=\frac{\mathbf{p}}{m},\quad\mathbf{\dot{p}}%
=-\bm{\nabla}V(\bm{q})-\zeta\mathbf{p}, \label{nh0}%
\end{equation}
(Eq.~(\ref{nh0})reads: $\quad
\mathbf{\dot{q}}_{k}=\mathbf{p}_{k}\mathbf{/}m_{k}, \:
\mathbf{\dot{p}}_{k}=-\bm{\nabla}_{\mathbf{q}_{k}}V(\bm{q})
-\zeta\mathbf{p}_{k},\: k=1,...,N \;) ;$ in what follows we use the
short notation above as it common in literature). Requiring that the
Liouville equation corresponding to (\ref{nh0}) has the canonical
distribution,
\[
\rho_{\infty}\propto \exp \left\{ -\beta \left[
\sum{\frac{\mathbf{p}^2}{2m}}+V({q})\right] \right\},
\]
(where $\beta\equiv1/(k_{\text{\textrm{B}}}T)$, $k_{
\text{\textrm{B}} }$ denotes the Boltzmann constant and $T$ is the
temperature), as a steady state solution, we obtain the following condition,%
\begin{equation}
\sum\frac{{\mathbf{p}}^{2}}{m}={\mathrm{d}N}k_{\text{\textrm{B}}}T. \label{c1}%
\end{equation}
It means that the total kinetic energy is constant and equal to 
equilibrium value according to the equipartition theorem. In other
word, with set (\ref{nh0}) and condition (\ref{c1}) we arrive at the
so-called Gaussian isokinetic thermostat \cite{T2}. This thermostat
does not extend system phase space $\mathcal{M}$. Nonholonomic
condition (\ref{c1}) provides for the temperature control and
requires nonconstant $\zeta$ (in contradiction with the original
supposition). The Gaussian isokinetic thermostat does not generate
the canonical statistics. To preserve in essential the structure of
(\ref{nh0}) and the corresponding Liouville equation with constant
$\zeta$ we need extend the phase space $\mathcal{M}$ and consider
$\zeta$ as an independent variable.

The Nos\'{e}-Hoover method is based on the idea of extended dynamics. When
variable $\zeta$ is endowed with its own equation of motion,%
\begin{equation}
\dot{\zeta}=g(\bm{q},\bm{p}), \label{g1}%
\end{equation}
then set (\ref{nh0}) and (\ref{g1}) represents an autonomous system,
and variable $\zeta$ simulates the thermostat in such a way that
(\ref{c1}) takes place only after time averaging. Thus the kinetic
energy is allowed to dynamically fluctuate around its equilibrium
value. More precisely, the Liouville equation corresponding to set
(\ref{nh0}) and (\ref{g1}) has a steady states solution of the form
\[
\rho_{\infty}\propto\exp\left\{  -\beta\left[
\sum{{\mathbf{p}}^{2}}/({2m)}+V(\bm{q})+\Phi(\zeta)\right]  \right\}
\]
only if
\begin{equation}
\Phi(\zeta)=\frac{1}{2}Q\zeta^{2},\quad g=\frac{1}{Q}\left(  \sum
\frac{{\mathbf{p}}^{2}}{m}-{\mathrm{d}N}k_{\text{\textrm{B}}}T\right)
,
\label{nh0g1}%
\end{equation}
where $Q$ is a constant. Thus the function $g$ is fixed up to a
constant multiplier. The parameter $Q\equiv
\mathrm{d}Nk_{\mathrm{B}}T\tau_{\mathrm{p}}^{2}$ appears as a
characteristic time scale $\tau_{\mathrm{p}}$. After the time
averaging,
\[
\overline{\left(  \cdots\right)  }=\lim\limits_{\Theta\rightarrow\infty}%
\frac{1}{\Theta}\int_{0}^{\Theta}dt\left(  \cdots\right),
\]
assuming the motion is bounded, Eq.~(\ref{g1}) leads to relation
\[
\overline{\left(  \sum\frac{{\mathbf{p}}^{2}}{m}\right)  }={\mathrm{d}N%
}k_{\text{\textrm{B}}}T
\]
that is in agreement with the equipartition theorem.

To conclude this section we note that the Nos\'{e}-Hoover extended
dynamics allows Hamiltonian reformulation. Consider balance of the
mechanical
work along trajectories of (\ref{nh0})-(\ref{g1}),%
\[
\sum-\bm{\nabla} V\cdot d\mathbf{q}=d\left(  \sum\frac{{\mathbf{p}}^{2}}{2m}%
+\frac{1}{2}Q\zeta^{2}\right)  +{\mathrm{d}N}k_{\text{B}}T\zeta dt.
\]
To obtain an exact differential equation, it is necessary to set%
\[
\zeta dt=d\lambda.
\]
In that case the following first integral is apparent,
\begin{equation}
I_{\text{NH}}=\sum\frac{{\mathbf{p}}^{2}}{2m}+V(\bm{q})+\frac{1}{2}Q\zeta
^{2}+{\mathrm{d}N}k_{\text{B}}T\lambda\;. \label{Inh}%
\end{equation}
Since the origin of the redundant variable $\lambda$ is arbitrary,
it is always possible for an arbitrary fixed trajectory to set
$I_{\text{NH}}=0$. Integral of motion (\ref{Inh}) is apparent
starting point for the Hamiltonian reformulation of the
Nos\'{e}-Hoover dynamics \cite{dettmann97,sc05}. The Hamiltonian
function has the form,%

\[
\mathcal{H}(\bm{q,}\lambda;\bm{u},\psi)=e^{-\lambda}{\sum
{\frac{\mathbf{u}{^{2}}}{{2m}}+{e^{\lambda}}V(\bm{q})}}+e^{-\lambda}%
\frac{1}{2Q}\psi^{2}+e^{\lambda}\mathrm{d}Nk_{\mathrm{B}}T\lambda,
\]
where the canonical variables,\ $\bm{u}=\{\mathbf{u_{k}}\}_{k=1}^N$
and $\psi$, are connected with the Nos\'{e}-Hoover dynamical
variables, $\bm{p}$ and $\zeta,$ by the relations
\[
\bm{u}={\exp(\lambda)}\bm{p},\quad \psi={\exp(\lambda)}Q\zeta.
\]
 Equations of motion,%
\[
\mathbf{\dot{q}}=\bm{\nabla}_{\mathbf{u}}\mathcal{H},\quad
\mathbf{\dot{u}}=-\bm{\nabla}_{\mathbf{q}}\mathcal{H},\quad
\dot{\lambda}={\nabla}_{\psi}\mathcal{H},\quad\dot{\psi
}=-{\nabla}_{\lambda}\mathcal{H},
\]
coincide with the Nos\'{e}-Hoover thermostat equations
(\ref{nh0})-(\ref{g1}) on the level set
\[
I_{\mathrm{NH}}=0.
\]

\subsection{The virial theorem and virial thermostat}
\label{sec:2.3}

The virial theorem is proved in classical as well as statistical
mechanics \cite{LL-M,LL-SM,Gal99}. This theorem has a great
generality since it does not require even ergodicity of motion. This
means that the temperature control below is valid when the
Nos\'{e}--Hoover thermostat is valid. There is a difference of kind
between the kinetic temperature and virial temperature control in
the time scale of thermostatting dynamics. The virial in connection
with the Nos\'{e}--Hoover thermostat first appears
in~\cite{hamilton90,hamilton93} in the context of the harmonic oscillator. In
this section we consider a more general case.

Consider a system of $N$\ particles as above. The quantity
$\mathcal{V}(\bm{q})$,
\[
\mathcal{V}(\bm{q})=\sum\mathbf{q}\cdot\bm{\nabla}V(\bm{q}),%
\]
defines the virial of the forces in the configuration $q$ (in short, the
virial). The virial theorem (Clausius) states the following relation of the
time averages,%
\begin{equation}
\overline{\left(  \sum\frac{{\mathbf{p}}^{2}}{m}\right)  }=\overline
{\mathcal{V}(\bm{q})}. \label{VirialT}%
\end{equation}
 The same relation is valid for the equilibrium averages in the canonical
ensemble as well.The immediate corollary of this relation,%
\begin{equation}
\overline{\mathcal{V}(\bm{q})}=\mathrm{d}Nk_{\text{\textrm{B}}}T, \label{v1}%
\end{equation}
clearly suggests the virial $\mathcal{V}(\bm{q})$ for the
temperature control in a thermostat similar to the Nos\'{e}-Hoover
one.

The virial $\mathcal{V}(\bm{q})$ of a system in a volume $v$ can be
expressed as sum of the virial of internal forces,
$\mathcal{V}_{int}(\bm{q})$, and the virial of external forces on
the boundary of the volume $v$, $\mathcal{V}_{ext}(\bm{q})$.
In result we arrive at the following relation,%
\begin{equation}
\overline{\mathcal{V}_{int}(\bm{q})}+\mathrm{d}Pv={\mathrm{d}N}k_{\text{\textrm{B}}}T,
\label{v2}%
\end{equation}
where $P$ is the pressure (the second theorem of Clausius; we refer
to \cite{Gal99,LL-SM} for details). In the case of free particles,
$\overline {\mathcal{V}_{int}(\bm{q})}=0$, and relation (\ref{v2})
is the ideal gas equation ($Pv=Nk_{\text{\textrm{B}}}T $). Relation
(\ref{v2}) is useful when the pressure control is required.

Analogues of the Gaussian isokinetic thermostat and the
Nos\'{e}-Hoover thermostat but under the temperature control
provided by the virial instead of the kinetic energy can be
formulated in the following manner. First, in the
same situation as in Section \ref{sec:2.2}, consider deformation
of the Hamiltonian system,%
\[
\mathbf{\dot{q}}=\frac{\mathbf{p}}{m},\quad\mathbf{\dot{p}}%
=-\bm{\nabla}V(\bm{q}),
\]
with a scalar parameter $\eta=const$ such that the Liouville
equation corresponding to the deformed dynamics has the canonical
distribution,
\[
\rho_{\infty}\propto\exp\left\{  -\beta\left[  \sum{{\mathbf{p}}^{2}}%
/({2m)}+V(\bm{q})\right]  \right\} ,
\]
as a steady state solution under condition 
\begin{equation}
\mathcal{V}(\bm{q})={\mathrm{d}N}k_{\text{\textrm{B}}}T. \label{c2}%
\end{equation}
Since the virial contains gradients of $V(\bm{q})$ the requirements
above define
the following deformed equations of motion,%
\begin{equation}
\mathbf{\dot{q}}=\frac{\mathbf{p}}{m}+\eta\mathbf{q},\quad\mathbf{\dot{p}%
}=-\bm{\nabla}V(\bm{q}). \label{vt0}%
\end{equation}
It can be verified by direct calculation that the Liouville equation
corresponding to (\ref{vt0}) has the canonical distribution as a
steady state solution only if (\ref{c2}) is valid. In other words,
the virial is a constant equal to equilibrium average value
(\ref{v1}). By analogy with the isokinetic thermostat we denote
(\ref{vt0}) together with condition (\ref{c2}) as the isovirial
thermostat. The isovirial thermostat does not extend system phase
space $\mathcal{M}$. Condition (\ref{c1}) provides for the
temperature control and requires nonconstant $\eta$ (in
contradiction with the original supposition). The isovirial
thermostat does not generate the canonical statistics. To preserve
in essential the structure of (\ref{vt0}) and the corresponding
Liouville equation with constant $\eta$ we need extend the phase
space $\mathcal{M}$ and consider $\eta$ as an independent variable.

Now we endow $\eta$ with its own equation of motion as in the Nos\'{e}--Hoover
dynamics and consider the following extended dynamics,%
\begin{equation}
\mathbf{\dot{q}}=\frac{\mathbf{p}}{m}+\eta\mathbf{q},\quad\mathbf{\dot{p}%
}=-\bm{\nabla}V(\bm{q}),\quad\dot{\eta}=h(\bm{q},\bm{p}), \label{vt1}%
\end{equation}
The Liouville equation corresponding to system (\ref{vt1}) has a
steady state solution of the form,
\[
\rho_{\infty}%
\propto\exp\left\{  -\beta\left[  \sum{{\mathbf{p}}^{2}}/({2m)}+V(\bm{q}%
)+\Psi(\eta)\right]  \right\},
\]
only if%
\begin{equation}
\Psi(\eta)=\frac{1}{2}Q\eta^{2},\quad h=\frac{1}{Q}\left(  {\mathrm{d}N%
}k_{\text{\textrm{B}}}T-\mathcal{V}(\bm{q})\right)  , \label{vt1h0}%
\end{equation}
where $Q$ is a constant. Thus function $h$ is fixed up to a constant
multiplier. Parameter $Q\equiv
\mathrm{d}Nk_{\mathrm{B}}T\tau_{\mathrm{q}}^{2}$ appears as a
characteristic time scale $\tau_{\mathrm{q}}$. It should be observed
here that time scales $\tau_{\mathrm{q}}$ and $\tau_{\mathrm{p}}$
(Section \ref{sec:2.2}) are possibly different.

Remark: The virial theorem has no relation to time scales
$\tau_{\mathrm{p}}$ (see below Eq.(\ref{nh0g1})) and
$\tau_{\mathrm{q}}$. Characteristic scales for time averaging of
left-hand side and right-hand side of (\ref{VirialT}) are possibly
different \ Since the temperature control based on the virial is
configurational we can \textit{a priori} expect that the
corresponding dynamics relates to slower processes then the
Nos\'{e}-Hoover dynamics. However this point requires a special
investigation, for example, to make a comparison of autocorrelation
functions of the Nos\'{e}-Hoover and the virial thermostatting
dynamics. We do not discuss this problem here.

Equation (\ref{vt1}) after time averaging
leads to the expected relation,%
\[
\overline{\mathcal{V}(\bm{q})}={\mathrm{d}N}k_{\text{\textrm{B}}}T.
\]

The virial thermostatting dynamics allows Hamiltonian reformulation. Consider
balance of the mechanical work along trajectories.%
\[
\sum-\bm{\nabla}V\cdot d\mathbf{q}=d(\sum\frac{\mathbf{p}^{2}}%
{2m}+\frac{1}{2}Q\eta^{2})-\mathrm{d}Nk_{\mathrm{B}}T\eta dt.
\]
To obtain an exact differential equation, it is necessary to set%
\[
\eta dt=d\mu.
\]
In that case, we obtain the following integral of motion,
\begin{equation}
I_{\text{V}}=\sum\frac{{\mathbf{p}}^{2}}{2m}+V(\bm{q})+\frac{1}{2}Q\eta
^{2}+{\mathrm{d}N}k_{\text{B}}T\mu\;. \label{I-V}%
\end{equation}
Since the origin of the redundant variable $\mu$ is arbitrary, it is
always possible for an arbitrary fixed trajectory to set
$I_{\text{V}}=0$. The integral of motion (\ref{I-V}) is apparent
starting point for the Hamiltonian reformulation of the virial
thermostatting dynamics (\ref{vt1}). The
Hamiltonian function has the form \cite{sc05},%
\[
\mathcal{H}(\bm{k},\lambda_{\eta};\bm{p},\varphi)=e^{-\mu}\sum
\frac{\mathbf{p}^{2}}{2m}+e^{-\mu}V(e^{\mu}\bm{k})+e^{\mu}\frac{1}%
{2Q}\varphi^{2}-e^{-\mu}\mathrm{d}Nk_{\mathbf{B}}T\mu,
\]
where the canonical variables,\ $\bm{k}=\{\mathbf{k}_i\}_{i=1}^N$
and $\varphi$ $,$ are connected with corresponding dynamical
variables, $\bm{q}$ and $\eta,$ by the relations,
\[
 \bm{k}=e^{\mu}\bm{q}, \quad
\varphi=e^{-\mu}Q\eta.
\]

Canonical equations of motion,%
\[
\mathbf{\dot{k}}=\bm{\nabla}_{\mathbf{p}}\mathcal{H},\quad
\mathbf{\dot{p}}=-\bm{\nabla}_{\mathbf{k}}\mathcal{H},\quad\dot{\mu
}={\nabla}_{\varphi}\mathcal{H},\quad\dot{\varphi}%
=-{\nabla}_{\mu}\mathcal{H},
\]
coincide with the virial thermostat equations (\ref{vt1}) on the
level set
\[
I_{\mathrm{V}}=0.
\]

To conclude this section we point out that the virial thermostat is
very suggestive of a configurational thermostat scheme.. Namely,
consider dynamical system (\ref{vt1}). Since equation for thermostat
variable $\eta$ (\ref{vt1h0}) does not explicitly include momenta
variables $p$ and the virial has rate of converging to equilibrium
value supposedly different from kinetic energy then it is possible
to postulate that momentum
variables are relaxed, similar to overdamped regime, to%
\[
\mathbf{p}=-\tau\bm{\nabla}V(\bm{q}),
\]
where $\tau=const$ is a relaxation time, and
arrive at the following configurational dynamics,%
\begin{equation}
\mathbf{\dot{q}}=-\tau \frac{1}{ m}\bm{\nabla}V(\bm{q})+\eta
\mathbf{q},\quad\dot{\eta}=\frac{1}{Q}\left(  {\mathrm{d}N}k_{\text{\textrm{B}%
}}T-\mathcal{V}(\bm{q})\right)  . \label{q_eta}%
\end{equation}

Since the Liouville equation corresponding to this system has under
certain condition (see Section \ref{sec:3}) the Boltzmann
distribution as a steady state solution we can consider
(\ref{q_eta}) as a first step toward a deterministic fully
configurational thermostat. This problem is considered in more
general context in Section \ref{sec:3}.

\subsection{Stimulated Nos\'{e}--Hoover dynamics}
\label{sec:2.4}

The Nos\'{e}-Hoover chain method \cite{chain} is often used in
practice, \textit{e.g}. \cite{mp04}. It is based on observation that
variable $\zeta$ generates the Gaussian statistics as well as
variables $\bm{p}$. This observation immediately suggests to
thermostat variable $\zeta$ by new thermostat variable $\zeta_{1}$
in a same manner and so on, and thus stimulate the equilibrium
statistics of the thermostat \cite{chain}. The Nos\'{e}-Hoover chain
of the full length of
$\mathrm{M}$ has the form,%

\[
\mathbf{\dot{q}}=\frac{{\mathbf{p}}}{m},\quad\mathbf{\dot{p}}%
=-\bm{\nabla}V(\bm{q})-\zeta{\mathbf{p},\quad}\dot{\zeta}=g-\zeta_{1}%
\zeta,\quad\dot{\zeta}_{i}=\frac{1}{Q_{i}}\left(  Q_{i-1}\zeta_{i-1}%
^{2}-k_{\mathrm{B}}T\right)  -\zeta_{i+1}\zeta_{i},
\]
where $i=1,...,\mathrm{M}$, $\zeta_{0}=\zeta$, $\zeta_{\mathrm{M}+1}\equiv0$;
$\left\{  Q_{i}\right\}  $\ are constant parameters. In that case%
\[
\rho_{\infty}\propto\exp\left\{  -\beta\left[  \sum\frac{{{\mathbf{p}}^{2}}%
}{{2m}}+V(\bm{q})+\frac{1}{2}Q\zeta^{2}+\sum_{i=1}^{\mathrm{M}}\frac{1}{2}%
Q_{i}\zeta_{i}^{2}\right]  \right\}  .
\]
The corresponding integral of motion has the following form,%
\[
I_{\text{NHchain}}=\sum\frac{{\mathbf{p}}^{2}}{2m}+V(\bm{q})+\frac{1}{2}Q\zeta
^{2}+\sum_{(i)}\frac{1}{2}Q_{i}\zeta_{i}^{2}+k_{\text{B}}T\lambda\;,
\]
where the redundant variable $\lambda$ satisfies equation%
\[
\dot{\lambda}={\mathrm{d}N}\zeta+\sum\limits_{(i)}\zeta_{i}.
\]

Since the Nos\'{e}-Hoover chain method is based on forcing the
Gaussian statistics of the thermostat variable(s) and thus speed up
the system to generate the equilibrium canonical statistics,  we can
propose a stochastic stimulation scheme as an alternative to the
chain method. The stochastic stimulation scheme becomes apparent
when it is considered that linear transformation of a Gaussian
random process is another Gaussian random process \cite{Gard04}. We
propose the following set of dynamical equations instead of the
Nos\'{e}-Hoover chain thermostatting dynamics,
\begin{equation}
\dot{{\mathbf{q}}}=\frac{{\mathbf{p}}}{m},\quad\dot{{\mathbf{p}}%
}=-\bm{\nabla}V(\bm{q})-\zeta{\mathbf{p}},\quad\quad\dot{\zeta}%
=g(\bm{p})-\gamma\zeta+f(t),\label{nh-sde}%
\end{equation}
where $f(t)$ is a generalized Gaussian random process (\textquotedblleft white
noise\textquotedblright), completely characterized by the first two
cumulants,
\[
\langle\,f(t)\,\rangle=0,\quad\langle\,f(t)f(t^{\prime})\,\rangle
=2D\delta(t-t^{\prime})\,.
\]
Thus the governing thermostatting equation (variable $\zeta$) is the
stochastic differential equation. In contrast to the Langevin method
\cite{Gard04} the only thermostat variable $\zeta$ is subject to
stochastic perturbation but the dynamical variables $\left(
q,p\right) $ are not directly stochastically perturbed. The
advantage over the chain method of the stochastic stimulation method
above consists in ensuring the ergodicity of the thermostat
\cite{GS72} .

The Liouville equation corresponding to (\ref{nh-sde}) after averaging over
all realizations of the random process $f(t)$ takes form of the Fokker--Planck
equation,
\begin{align}
\frac{{\partial\rho}}{{\partial t}}=  & \sum\left[  -\frac{{\mathbf{p}}}%
{m}\cdot\bm{\nabla}_{{{\mathbf{q}}}}{\rho}+\bm{\nabla
}_{{{\mathbf{q}}}}{V}\cdot\bm{\nabla}_{{{\mathbf{p}}}}{\rho}%
+\zeta\bm{\nabla}_{{{\mathbf{p}}}}\cdot\left(  {{\mathbf{p}}\rho
}\right)  \right] \nonumber \\
& -g\left(  p\right)  \frac{{\partial\rho}}{{\partial\zeta}}+\gamma
\frac{\partial}{{\partial\zeta}}\left(  {\zeta\rho}\right)  -D\frac
{\partial^{2}\rho}{{\partial\zeta^{2}}}\,.\label{FPL}
\end{align}
It only remains to prove that this equation has the steady state
solution $\rho_{\infty}$,
\[
\rho_{\infty}\varpropto\exp\left\{  -\beta\left[  \sum\frac{{{\mathbf{p}}^{2}%
}}{2m}+V(\bm{q})+\frac{1}{2}Q\zeta^{2}\right]  \right\}  \;.
\]
Substitution of $\rho_{\infty}$ into equation (\ref{FPL}) leads to the
following conditions,
\begin{align*}
D &  =\frac{\gamma}{Q}k_{\text{B}}T\;,\\
g(\bm{p}) &  =\frac{1}{Q}\left[  {\sum{\frac{{{\mathbf{p}}^{2}}}{m}-{\mathrm{d}N%
}k_{\text{B}}T}}\right]  \;.
\end{align*}
The last expression for the function $g$ is the same as
(\ref{nh0g1}) for the analogous function in the Nos\'{e}--Hoover
chain method.  The first equation is an analogue of the
fluctuation-dissipation relation.

\section{Configurational thermostat}
\label{sec:3}

It is reasonable, in the spirit of deterministic thermostat methods, to
conjecture that it is possible to use the relaxation time $\tau$ for
thermostatting configurational degrees of freedom when momentum variables are
still relaxed in their local equilibrium state. Of course, in this case the
sign of $\tau$ is not fixed and $V$ loses its meaning as a Lyapunov function.
In a sense, it means that time can go back as well as forward.

First, consider the simple dynamical equations for $\mathrm{d}N$
degrees of freedom,
\begin{equation}
\label{iso-S}m\mathbf{\dot{q}}=-\tau{\bm{\nabla}} V(\bm{q}),
\end{equation}
where $\tau$ is a constant, but by analogy with Nos\'e-Hoover will
be endowed with its own equation of motion below. Short notation as
in Section \ref{sec:2} is used in (\ref{iso-S}) and in what follows;
(\ref{iso-S}) reads as
$m_{k}\mathbf{\dot{q}}_{k}=-\tau{\bm{\nabla}}_{\mathbf{q}_{k}}V(\bm{q}),
k=1,...,N$. By the change of variables,
$\mathbf{x}=\sqrt{m}\mathbf{q}$, it is possible to exclude all
masses from the formulae in what follows, but we prefer to save the
physical notation. Requiring that the corresponding Liouville
equation has the Boltzmann distribution,
$\rho_{\infty}\propto\exp\left[  -V(\bm{q})/(k_{\mathrm{B}}T)\right]
$, as a steady state solution, we arrive at the condition that
involves the temperature in the dynamics,
\begin{equation}
\label{T}\sum\frac{1}{m}\left[  \Delta
V(\bm{q})-\frac{1}{k_{\mathrm{B}}T}\left( \bm{\nabla}
V(\bm{q})\right) ^{2}\right]  =0.
\end{equation}
After time averaging, Eq.~(\ref{T}) appears as the definition of the
recently introduced so-called configurational temperature
\cite{rugh97,evans00}. Currently it is used in molecular dynamics
simulations \cite{delh04,delh04PE}. In a more general context, in the case of
a presupposed anisotropy in the system, let us assume that $\tau$ in
(\ref{iso-S}) is a matrix, $\tau\to\tens{\Gamma}$. Then the dynamics
take the form, $m\mathbf{\dot{q}}=-\tens{\Gamma}\bm{\nabla}
V(\bm{q})$, and the condition that involves the temperature in the
dynamics is
\[
\sum\frac{1}{m}\left[  (\bm{\nabla},\tens{\Gamma} \bm{\nabla})
V(\bm{q})-\frac {1}{k_{\mathrm{B}}T}\left(  \bm{\nabla}
V(\bm{q}),\tens{\Gamma} \bm{\nabla} V(\bm{q}) \right)  \right]  =0.
\]
Note that this condition involves the presupposed time scales in the dynamical
temperature control. Conventional Nos\'e-Hoover methods do not allow such a
generalisation. In what follows we also consider $\tau$ as a scalar. On the
other hand, it is useful to keep in mind the possibility of the generalisation.

We now attempt to generate statistics as in the Nos\'e-Hoover scheme
by making $\tau$ an independent variable in~(\ref{iso-S}). It is
easily seen that this is too simple. At an equilibrium point
$\bm{\nabla} V=0$, that is, all forces are zero, the evolution comes
to a halt and no longer fluctuates, irrespective of the time
dependence of $\tau$. For initial conditions with nonzero
forces~(\ref{iso-S}) after a (positive or negative) change of time
variable, it is a gradient flow as defined in~\cite{KH}, and it is
easy to show that all trajectories move along paths in $q$ with
equilibrium points at either end. In short, the system is not
ergodic. Note also that~(\ref{T}) is singular when $\bm{\nabla}
V=0$.

The way to overcome this difficulty is suggested by the Smoluchowski
stochastic equation. In this equation the ergodic motion is ensured
by the random forcing. Hence, we need to add a deterministic
analogue of the random force term in~(\ref{Smoluchowski}). Let us
consider, instead of (\ref{iso-S}), the dynamical equations,
\begin{equation}
m\mathbf{\dot{q}}=-\tau\bm{\nabla}V(\bm{q})+\bm{\xi}, \label{xi}%
\end{equation}
where $\bm{\xi}$ are constant vectors, but they will be endowed with
their own equation of motion below. Requiring that the Liouville
equation corresponding to (\ref{xi}) has the Boltzmann distribution
as a steady state solution, together with temperature control
condition (\ref{T}) we arrive at the condition
\begin{equation}\label{fxi}
    \sum\frac{1}{m}\bm{\xi}\cdot\bm{\nabla}V(\bm{q})=0.
\end{equation}
To detail the nature of vectors $\bm{\xi}$, three principal cases
are possible: (a) All $\bm{\xi}=\{\bm{\xi}_{i}\}_{i=1}^{N}$, where
$N$ is number of particle in the system, can be varied
independently; (b) All $\bm{\xi}_{i}=\bm{\xi}$ are varied
identically; (c) There exist a preferred direction, $\mathbf{e}$,
where $\mathbf{e}$ is a constant unit vector of physical
(\textit{i.e.} three dimensional) space, and only one variable,
$\xi$, $\bm{\xi}=\xi\mathbf{e}$, is varied.  Varying $\bm{\xi}$ in
(\ref{fxi}) according to cases (a), (b) and (c) we correspondingly
obtain the following particular  conditions,
\begin{equation}
(a)\bm{\nabla}V(\bm{q})=0,\quad(b)\sum\frac{1}{m}\bm{\nabla}
V(\bm{q})=0,\quad
(c)\sum\frac{1}{m}\mathbf{e}\cdot\bm{\nabla}V(\bm{q})=0. \label{ffxi}%
\end{equation}
These conditions do not involve temperature but the thermalized
forces acting in the system. All of them, as well as a their
combination, are candidates for simulating the deterministic
analogue of the random force, chosen according to the problem under
consideration. For example, in respect of the Peyrard-Bishop
dynamical model of the DNA molecule \cite{mp04}, case (c) appears to
be appropriate.

The physical sense of the conditions above are the following.

Case (a): the force acting on a particle in the system equals zero
(static equilibrium of forces). In that case dynamical equations
(\ref{xi}) degenerate to triviality, $\bm{\xi}=0$ and the
temperature control condition takes a sense only if the temperature
$T=0$. But when $\bm{\xi}$ are endowed with their own equations of
motion and conditions (a) take place only after time averaging then
dynamics (\ref{xi}) is robust and $\bm{\xi}$ provide a shaking of
the system around the configuration of the mechanical equilibrium of
the system.

Case (b): the static equilibrium of forces is not required but the
total force acting on the system equals zero (stability of the
system). In that case when $\bm{\xi}$ is endowed with its own
equation of motion, it provides a shaking of the system around its
center of inertia.

Case (c): stability of the system in the direction $\mathbf{e}$.
When ${\xi}$ is endowed with its own equation of motion, it provides
a shaking of the system along the direction $\mathbf{e}$.

In all cases the time scale of such a shaking is still a parameter
of the theory.

It is practical to remark, in consideration of the virial thermostat
scheme of Section \ref{sec:2.3}, that a thermostatting dynamics more
general than (\ref{xi}) is possible. Consider dynamical equations of
the form
\begin{equation}
\label{VSdyn}m\mathbf{\dot{q}}=-\tau\bm{\nabla} V(\bm{q})+\eta m\mathbf{q}%
+\bm{\xi},
\end{equation}
where the term $\eta m \mathbf{q}$ is suggested by the virial
thermostatting scheme. Requiring that the Liouville equation
corresponding (\ref{VSdyn}) has the Boltzmann distribution as a
steady state solution we obtain together with (\ref{T}) and
\ref{fxi}) the following condition on the virial function,
\begin{equation}
\label{V}\mathrm{d}Nk_{\mathrm{B}}T-\sum\mathbf{q}\cdot\bm{\nabla}
V=0.
\end{equation}
In this case a double temperature control is provided. The $\eta$
term is not a mandatory temperature control for our configurational
thermostat. We could consider only the $\tau$ term by a trivial
modification of~(\ref{VSdyn}) and subsequent equations but consider
a more general equations of the form (\ref{VSdyn}). The reason is
that the virial function involves the pressure in the dynamics and
this is important for biologically oriented models. For the sake of
definiteness and keeping in mind a future application we here fix
case (c). When $\xi$ is endowed with its own equation of motion, it
provides a shaking of the system along the direction $\mathbf{e}$.

Now we have three parameters which we group together as a 3-vector
$\bm{\alpha}=(\tau,\eta,\xi)^{\mathrm{T}}$. As in~(\ref{T}), (\ref{fxi}) and
(\ref{V}) we find a stationary solution of the Liouville equation of the form
\[
\rho_{\infty}\propto\exp\left[  -(V(\bm{q})+\bm{\alpha}^{\mathrm{T}}%
\tens{Q}\bm{\alpha}/2)/(k_{\mathrm{B}}T)\right]
\]
where $\tens{Q}$ is a positive definite real symmetric matrix. It
is, of course, not possible to justify consistently that parameters
$\bm{\alpha}$ have a Gaussian distribution at equilibrium. It is
just a reasonable assumption that leads to the simplest form of
self-consistent thermostat dynamics. Intuitively, we can justify the
Gaussian character of the parameters at equilibrium by the limiting
theorems of probability theory. Note that when multiple thermostats
are present, they are usually assumed to be uncoupled (diagonal
$\tens{Q}$); mathematically it is not required and we do not feel
that this is physically necessary, thus we consider that the more
general case of coupled thermostats may be useful. For comparison
with the uncoupled case we define
$\tilde{\tens{Q}}=diag(Q_{\tau\tau},Q_{\eta\eta},Q_{\xi\xi})$ to be
the matrix with only the diagonal components of $\tens{Q}$. Instead
of Equations (\ref{T}), (\ref{ffxi}) and (\ref{V}) we obtain
\begin{equation}
\label{VStemp}\tilde{\tens{Q}}\bm{g} \equiv\left(
\begin{array}
[c]{c}%
\sum\frac{1}{m}\left[  (\bm{\nabla} V)^{2}-k_{\mathrm{B}}T\Delta V\right] \\
\mathrm{d}Nk_{\mathrm{B}}T-\sum\mathbf{q}\cdot\bm{\nabla} V\\
-\sum\frac{1}{m}\mathbf{e}\cdot\bm{\nabla} V
\end{array}
\right)  =0,
\end{equation}
which now defines $\bm{g}$. It should be remembered (see Sections
\ref{sec:2.2} and \ref{sec:2.3}) that conditions
$\tilde{\tens{Q}}\bm{g}=0$ are inconsistent with the initial
supposition of $\bm{\alpha}=const$.

Let us realize the main conjecture the configurational thermostat
scheme and allow the components of $\bm{\alpha}$ to fluctuate so
that~(\ref{VStemp}) holds only after time averaging. Thus we extend
the original phase space $\mathcal{M}$. We write
\begin{equation}
\label{alphadot}\dot{\bm{\alpha}}=\bm{G}%
\end{equation}
where $\bm{G}$ is as yet an undetermined vector of functions. Now
requiring the same condition for the solution of the Liouville
equation corresponding to system (\ref{VSdyn}) and (\ref{alphadot}),
we find that the only solution is
\begin{equation}
\label{G}\bm{G}=\tens{Q}^{-1}\tilde{\tens{Q}}\bm{g}.
\end{equation}
Thus the only undetermined parameters of our thermostatting scheme
are the components of the positive definite real symmetric matrix
$\tens{Q}$, and in the uncoupled case $\tilde{\tens{Q}}=\tens{Q}$ we
have $\bm{G}=\bm{g}$. In this uncoupled case dynamics takes the
simple form,
\begin{eqnarray*}
    m_{k}\mathbf{\dot{q}}_{k} &=& -\tau\bm{\nabla}_{\mathbf{q}_{k}} V(\bm{q})
    +\eta m_{k}\mathbf{q}_{k}+\bm{\xi},\\
    \dot{\tau} &=& \frac{1}{Q_{\tau\tau}}\sum_{k=1}^{N}\frac{1}{m_{k}}\left[
    (\bm{\nabla}_{\mathbf{q}_{k}} V)^{2}-k_{\mathrm{B}}T\Delta_{\mathbf{q}_{k}} V\right], \\
    \dot{\eta} &=& \frac{1}{Q_{\eta\eta}}(\mathrm{d}Nk_{\mathrm{B}}T
    -\sum_{k=1}^{N}\mathbf{q_{k}}\cdot\bm{\nabla}_{\mathbf{q}_{k}} V), \\
    \bm{\dot{\xi}} &=&
    -\frac{1}{Q_{\xi\xi}}\sum_{k=1}^{N}\frac{1}{m_{k}}\mathbf{e}\cdot\bm{\nabla}_{\mathbf{q}_{k}}V .
\end{eqnarray*}
Parameters $Q_{\tau\tau},Q_{\eta\eta},Q_{\xi\xi}$ define time scales
that are in general different.

We can now ask whether the addition of new variables $\eta$ and/or
$\xi$ will remove the lack of ergodicity implied by the potential
flow argument applying to~(\ref{iso-S}). A partial answer is
provided by the Frobenius theorem of differential
geometry~\cite{Lang}, which in our case states that an integral
surface exists (hence the dynamics are definitely not ergodic) if a
vector space containing the terms in the equation for $\mathbf{q}$
but smaller than the full phase space is closed under Lie brackets.
For realistic potentials (not the harmonic oscillator) this is very
unlikely since multiple derivatives of $V$ are almost always
linearly independent. If the theorem does not apply we are in the
same situation as for nonthermostatted nonintegrable many particle
systems, which are often assumed to be ergodic, at least for
practical purposes.

Since under the transformation $t\rightarrow-t, \bm{\alpha}\rightarrow
-\bm{\alpha}$ the equations of motion (\ref{VSdyn}) and (\ref{alphadot}) are
still unchanged, they are time reversible.

To find a mechanically important integral of motion of system (\ref{VSdyn}),
we need to add a redundant variable. Indeed, consider the balance of the
mechanical work along trajectories of Eqs.~(\ref{VSdyn}) and (\ref{alphadot}),%
\[
\sum-\bm{\nabla} V\cdot d \mathbf{q}= d (\bm{\alpha}^{\mathrm{T}%
}\tens{Q}\bm{\alpha}/2)+k_{\mathrm{B}}T\left(  \sum\frac{\Delta V}{m}%
\tau-\mathrm{d}N\eta\right)  d t.
\]
To obtain an exact differential equation, it is necessary to set%
\[
\left(  \sum\frac{\Delta V}{m}\tau-\mathrm{d}N\eta\right)  d t= d
\theta.
\]
In that case, the following integral of motion is apparent,%
\[
I_{\mathrm{S}}=V(  \bm{q})  +\bm{\alpha}^{\mathrm{T}}%
\tens{Q}\bm{\alpha}/2+k_{\mathrm{B}}T\theta.
\]
Since the origin of the redundant variable $\theta$ is arbitrary, it
is always possible for an arbitrary fixed trajectory to set
$I_{\mathrm{S}} =0$. This integral of motion is apparent control
parameter in numerical simulations. Besides, it clearly relates to
the equilibrium distribution $\rho_{\infty}$ and thus can be
considered as a first step to reformulation of Eqs.~(\ref{VSdyn})
and (\ref{alphadot}) in terms of a free energy functional. Recall
that the corresponding Nos\'{e}-Hoover integral of motion, Section
\ref{sec:2.2}, as well as the virial scheme integral of motion,
Section \ref{sec:2.3}, both are given by their Hamiltonian - no
Hamiltonian is possible here since the momentum does not appear
explicitly.

\section{Stimulated configurational thermostats}

\label{sec:4}

Since the configurational dynamics above result in the Gaussian equilibrium
fluctuation of thermostat variables, the latter admits reinforcing by a chain
of equations analogous to the Nose-Hoover chain thermostat \cite{chain}. The
chain method consists in including a subsidiary sequence of dynamical
variables, $\{\bm{\alpha}_{i}\}$, into a thermostat scheme such that
asymptotically, in the equilibrium distribution, they are independent Gaussian
variables,
\[
\rho_{\infty}\propto\exp[-(V(\bm{q})+\frac{1}{2}\bm{\alpha}^{\mathrm{T}%
}\tens{Q}\bm{\alpha}+\sum\nolimits_{(i)}\frac{1}{2}\bm{\alpha}_{i}%
^{\mathrm{T}}\tens{Q}_{i}\bm{\alpha}_{i})/(k_{\mathrm{B}}T)].
\]
The corresponding dynamics are not unique. We have obtained a clear method for
generalizing the chain, but since this does not directly relate to our main
topic we do not discuss the details. Instead we cite the example of the chain
rule that has been used in our test simulation,
\begin{align}
\label{chain}m{\mathbf{\dot{q}}}  &  =-\tau\bm{\nabla}V(\bm{q})+\xi\bm{e}%
,\;\dot{\tau}=g_{\tau}+\tau_{1}\tau,\;\nonumber\\
\dot{\tau}_{i}  &  =\frac{1}{Q_{\tau_{i}}}\left(  k_{\text{\textrm{B}}%
}T-Q_{\tau_{i-1}}\tau_{i-1}^{2}\right)  +\tau_{i+1}\tau_{i},\;\dot{\xi}%
=g_{\xi},
\end{align}
where $i=1,\ldots,\mathrm{M},:\tau_{0}=\tau,:\tau_{\mathrm{M}+1}\equiv0$. It
is a simple chain of total length $\mathrm{M}$.

It should also be noted that in spite of its popularity, the effectiveness of
the chain method for computing non-equilibrium properties has been questioned
\cite{branka00,T4}.

It is possible to stimulate the Gaussian fluctuation of the
thermostat variables by the process of Brownian motion. This scheme
has the advantage of ensuring the ergodicity property. The
stimulation, similar to the chain one, may be done in a few ways,
applying to one or more of the variables $\tau$, $\eta$ and $\xi$
(see Section (\ref{sec:2.4}) for the prototype). In general we have
\begin{align}
\label{stst}m\mathbf{\dot{q}} =-\tau\bm{\nabla} V(\bm{q})+\eta m\mathbf{q}%
+\xi\mathbf{e}
,\quad\dot{\bm{\alpha}}=\bm{G}-\tens{\Lambda}\bm{\alpha}+\sqrt
{2\tens{D}}\bm{f}(t) ,
\end{align}
where now $\tens{\Lambda}$ and $\sqrt{2\bm{\mathrm{D}}}$ are
positive definite real symmetric matrices, and $\bm{f}(t)$ is a
vector of independent white noise components. The Liouville equation
corresponding to (\ref{stst}), averaged over all the realizations of
$\bm{f}(t)$, has the form of the Fokker-Planck equation (similar to
(\ref{FPL})) and the Boltzmann distribution as a steady state
solution only if
\[
{k_{\mathrm{B}}T}\,\tens{\Lambda}=\tens{D}\tens{Q}.
\]
We consider that the most physical case is when the noise is used
only for temperature control, that is, for $\tau$ and $\eta$ only.
We do not establish extreme generality here because our main aim is
the presentation of the idea of the \emph{deterministic} fully
configurational thermostat. The effectiveness of Eqs.~(\ref{stst})
for systems far from equilibrium is not clear.
\begin{figure}[ptb]
\includegraphics[clip,width=27pc]{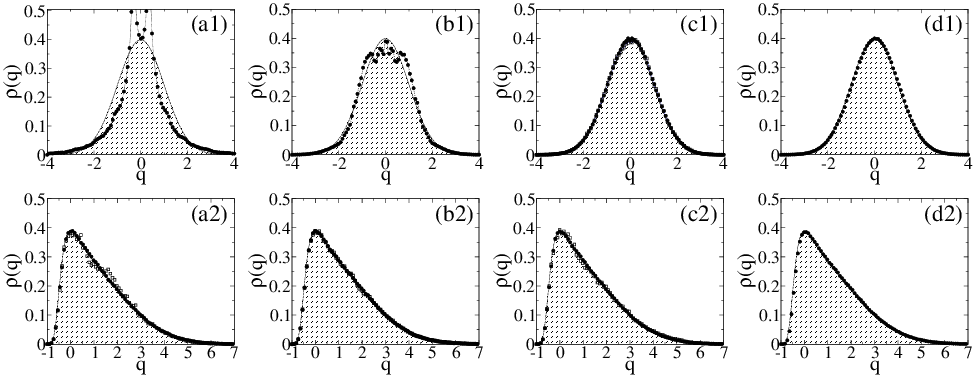}%
\caption{Probability distributions of position variable (shown on
background of exact analytical distribution) of the harmonic (1) and
Morse (2) oscillator. Probability densities are calculated as
normalized sojourn distributions. Correspondingly, thermostat is:
(a1)-(a2) non-stimulated and uncorrelated (Eq.~(\ref{VSdyn}),
$\dot\tau=g_{\tau}, \eta\equiv0, \dot \xi=g_{\xi}$); (b1)-(b2)
non-stimulated and uncorrelated but under double temperature control
(Eq.~(\ref{VSdyn}), $\dot\tau=g_{\tau}, \dot\eta=g_{\eta},
\dot\xi=g_{\xi}$); (c1)-(c2) uncorrelated but stimulated by the
chain rule (Eq.~(\ref{chain}), $\mathrm{M}=1$); (d1)-(d2) stimulated
by stochastic process (Eq.~(\ref{stst}), $\eta\equiv0, D=1$). All
simulations are performed at $Q_{\tau}=Q_{\xi}=1$, $Q_{\eta}=0.1$
for $t=10^{4}$ (squares) and
$t=10^{6}$ (black circles). }\label{fig1}%
\end{figure}

\section{Test numerical simulations}
\label{sec:5}

The harmonic oscillator is both a simple and an important physical
system. At the same time, it reveals the ergodicity problem in the
canonical ensemble simulation. For this reason, it is important to
test the configurational thermostat method capable of generating the
Boltzmann distribution for a single harmonic oscillator in one
dimension, $V(\bm{q})=q^{2}/2 $. Then it is reasonable to simulate
another one-dimensional system, `good' from the point of view of the
Frobenius theorem, and to compare results. We choose for this
purpose the Morse oscillator,
$V(\bm{q})=V_{0}(1-\exp(-aq))^{2}+\mathrm{k}q^{2}/2$.
Simulations are performed using global parameters $m=1$ and $k_{\mathrm{B}%
}T=1$, and the Morse potential parameters $V_{0}=0.25$, $a=2$, $\mathrm{k}%
=0.25$. Figure~\ref{fig1} shows the probability distribution of the
position variable calculated with the four simplest configurational
thermostats. Note the effectiveness of the double temperature
control. The ability of the configurational thermostat to reproduce
the correct distribution function, $\rho(\bm{q})$, even with
absolute minimum of this thermostat capacity, demonstrates its great
potential for application.

\section{Conclusion}
\label{sec:6}

An innovative constant temperature thermostat, the configurational
thermostat, exclusively involving dynamics of the configurational
variables has been introduced. It poses the general problem of the
derivation of a thermostatting dynamics for slow dynamical variables
and outlines at least one way of the solution. For practical
purposes, the new thermostatting scheme can easily be combined with
complementary temperature control via dynamical fluctuation of the
virial function. This combination helps to enhance the efficiency of
the thermostat temperature control as well as to implement pressure
control into the dynamics. It is also relevant to emphasise the
appearance of the dynamically fluctuated forces and correlated
temperature control in the presented thermostatting scheme. We
finally remark that the proposed method is applicable to
thermostatting such macro/meso-scale models as the
reaction-diffusion dynamical equations.

\newpage

\begin{acknowledgements}
AS is grateful for support from the University of Bristol and the
University of Dundee, from CCP5, and from the Royal Society
(London).
\end{acknowledgements}

\end{document}